\documentclass[prd,12pt]{revtex4}
\usepackage{color,slashed}
\usepackage{graphicx}
\newcommand\TR[1]{\langle{#1}\rangle}
\newcommand\Higgs{{\it The Boson }}
\begin{document}

\title{Combined Analysis of the CP Properties of The Higgs Boson in Effective Higgs Lagrangian}
\author{Rong Li\footnote{E-mail:rongliphy@mail.xjtu.edu.cn}, 
	Ying Zhang\footnote{E-mail:hepzhy@mail.xjtu.edu.cn. Corresponding author. }
}
\address{School of Science, Xi'an Jiaotong University, Xi'an, 710049, China}
\begin{abstract}
CP violation effects of the Higgs stem from not only CP-violation interactions but also an ambiguous defined CP state. The two CPV sources are coherently studied based on an effective Higgs Lagrangian.
The constraints from unitarity limits for $WW$ and $ZZ$ scatterings are proposed to restrict Higgs couplings to weak gauge bosons. Five interesting cases are classified in terms of CPV sources to investigate the Higgs CP properties. The allowed ranges are shown from fitting results to the signal strengths of the Higgs measured by ATLAS and CMS.

\bigskip



\end{abstract}
\maketitle

%%%%%%%%%%
\section{Motivation}
Although much work has been done on \Higgs, found in 2012 at the LHC with a mass of about $125$~GeV \cite{HiggsFound}, there is no definite answer yet whether it is the Standard Model (SM) Higgs boson or not. CP property and self-interaction are two important check points. Compared with the latter, the former is more easy to examine in present and coming experiments. 
Differing from the CP-even state in SM, \Higgs can be explained as an ambiguous CP defined state, like a mixing of a CP-even state and a CP-odd state in the two Higgs doublet model (2HDM) \cite{2HDM}.
It suggests a kind of CP violation (CPV) effect. The CP mixing angle has been constrained by the Higgs experiment signal in much of the literature \cite{HiggsCP,reRatio}. Additionally, the CPV effect can also arise from another source, namely, CPV interactions beyond the SM. 
In Ref.\cite{HVVexp}, CPV couplings in $HVV$ vertices (with $V=Z,W$) are required to describe the general tensor structures appearing in high energy experiments. Other CPV interactions with defined CP parity Higgs are often be investigated in terms of high dimension operators in \cite{ChangPRD,GavelaJHEP}.
Nevertheless, two kinds of CPV effect, ambiguous CP defined state and CPV interactions, can correct Higgs production and decay channels measured by ATLAS and CMS, individually and coherently. The two must also be analyzed together. Indeed, there is no reason to neglect one or other. Differing from articles addressing a single CPV source, the purpose of this paper to present a combined analysis of the two kinds of CPV effect in observations of \Higgs, thereby providing a better means to discriminate between the SM Higgs and beyond.

To start a general investigation, we first review a Higgs electroweak chiral effective theory in Sec.\ref{sec.Lag}. The proposal in Ref.\cite{LMWangHiggs} is a development with a scalar Higgs from the electroweak chiral Lagrangian in Ref.\cite{AppelquistEWCL}.
As electroweak symmetry can be broken according to many models that go beyond the SM, such as the strong coupled Higgs model, where the Lagrangian adopts a nonlinear-realized Goldstone boson to describe electroweak symmetry breaking. The physical Higgs after electroweak symmetry broken is only treated as a singlet scalar. We add CPV Higgs interactions to the Lagrangian to complete it in $p^4$ order. The effective Higgs Lagrangian provides a larger parameter space to investigate CP properties of \Higgs.
%%%
In Sec.\ref{sec.HVV}, we investigate the Higgs interactions with weak gauge bosons. These couplings provide not only CP conserved vertexes but also CPV form. On the other hand, they are consistently associated with Higgs productions and decay at the LHC. The relation between the $HVV$ tensor structure and low-energy coefficients of the Lagrangian is studied. Theoretical constraints are also proposed from unitarity limits in $WW/ZZ$ scatterings.
%%%
In Sec.\ref{sec.phenom}, we classify five interesting cases according to the CPV source. Five classes are identified and we calculate the Higgs decay width into the different channels of each class. To avoid complexity in the calculations of Higgs production, we choose the sum of signal strengths $\mu_{ref}=\sum_X\mu_X$ as a benchmark to define $H\rightarrow XX$ relative ratios of branching ratios $D_x\equiv \mu_x/\mu_{ref}$.
%%%
Based on this study, we fit the parameters in Higgs effective theory to the experimental signal strength at $\sqrt{s}=8$~TeV and in Sec.\ref{sec.fit} give the optimal parameter values and ranges for the different cases. Finally, a short summary is given in Sec.\ref{sec.summary}.
%%%%%%%%%%%
\section{Nonlinear-realized effective Higgs Lagrangian}\label{sec.Lag}
In the section, we review the construction of a nonlinear-realized effective Higgs Lagrangian presented in \cite{LMWangHiggs} and generalize it to include CPV terms.

The Goldstone bosons is expressed as unitary unimodular matrices $U$ that are associated with the cosets of $SU(2)_L\otimes U(1)_Y/U(1)_{em}$.
A physical Higgs after EW symmetry broken is labeled by a singlet scalar $h$.
The covariant derivatives of the Goldstone fields are written
\begin{eqnarray}
D_{\mu}U=\partial_{\mu}U+ig\frac{\tau^a}{2}W^a_{\mu}U-igU\frac{\tau_3}{2}B_{\mu}.
\end{eqnarray}
Here, $W_{\mu}^a$ ($a=1,2,3$) and $B_{\mu}$ are $SU(2)_L$ and $U(1)_Y$ gauge fields, respectively.
We adopt $SU(2)_L$ covariant building blocks \cite{AppelquistEWCL}
\begin{eqnarray}
T_3=U^\dag\tau_3U,~~~V_\mu=(D_\mu U)U^\dag,
\end{eqnarray}
to express the effective Higgs Lagrangian. As $h$ is a singlet scalar, it couples to an effective field operator $\mathcal{O}$ in any way
\begin{eqnarray}
\left(1+\sum_{i=1}^Nc^i\frac{h^i}{\Lambda^i}\right)\cdot\mathcal{O}.
\label{eq.HiggsFunction}
\end{eqnarray}
The high power of the Higgs field implies that a strongly coupled case appearing at energy scale $\Lambda$ has been recovered. The factor including the Higgs field in Eq.\ref{eq.HiggsFunction} can be integrated into a random function for $h$, i.e., the Higgs field $h$ is counted as order $p^0$. The lowest order of the chiral Lagrangian is just that of the Higgs potential
\begin{eqnarray}
\mathcal{L}_0&=&-V(h)
\end{eqnarray}
The details of $V(h)$ depend on the nature of electroweak symmetry breaking. In general, the function can be expressed as a power series in the Higgs field $V(h)=\sum_{i=2}^\infty \lambda_i h^i$ that includes the mass term (for $i=2$) and all possible self-interactions. The next-to-leading order is $p^2$ for this Lagrangian,
\begin{eqnarray}
\mathcal{L}_2&=&-\frac{1}{4}f^2\TR{{V}_\mu{V}^\mu}
+\frac{1}{4}\beta f^2\TR{T{V}_\mu}^2+\frac{1}{2}(\partial_\mu h)^2
\label{eq.Lagp2}
\end{eqnarray}
in which $f$ takes the scale of spontaneous breaking for electroweak symmetry and $\langle~\rangle$ denotes trace in the weak isospin space. The first term is the Lagrangian of the nonlinear sigma model. The second corresponds to leading-order custodial symmetry-violating interactions. Because $\beta$ depends on the Higgs field $h$, it involves interactions with the EW gauge bosons. When neglecting the non-standard EW rotations between $W^3_\mu$ and $B_\mu$, the term yields the $H^NZZ$ vertex with $N$ Higgs legs. The last term in Eq.\ref{eq.Lagp2} is the Higgs kinetic term.

The Lagrangian at $p^4$ order can be divided into three contributions,
\begin{eqnarray}
\mathcal{L}_4=\mathcal{L}_K+\mathcal{L}_B+\mathcal{L}_H
\end{eqnarray}
with kinetic term $\mathcal{L}_K$, interaction terms without derivatives of the Higgs $\mathcal{L}_B$, and those with derivatives of the Higgs $\mathcal{L}_H$. The kinetic term can be expressed as
\begin{eqnarray}
\mathcal{L}_K=-\frac{1}{2}\TR{W_{\mu\nu}W^{\mu\nu}}-\frac{1}{4}B_{\mu\nu}B^{\mu\nu}.
\label{LK}
\end{eqnarray}
$\mathcal{L}_{B}$ has a similar form to the electroweak chiral Lagrangian in \cite{AppelquistEWCL}
\begin{eqnarray}
\mathcal{L}_B
&=&\frac{1}{2}\alpha^{(0)}_1gg'B_{\mu\nu}\TR{TW^{\mu\nu}}
+\frac{i}{2}\alpha^{(0)}_2g'B_{\mu\nu}\TR{T[{V}^\mu,{V}^\nu]}
+i\alpha^{(0)}_3g\TR{W^{\mu\nu}[{V}^\mu,{V}^\nu]}\nonumber\\
&&+\alpha^{(0)}_4\TR{{V}_\mu{V}_\nu}^2
+\alpha^{(0)}_5\TR{{V}_\mu{V}^\mu}^2
+\alpha^{(0)}_6\TR{{V}_\mu{V}_\nu}\TR{T{V}^\mu}\TR{T{V}^\nu}\nonumber\\
&&+\alpha^{(0)}_7\TR{{V}_\mu{V}^\mu}\TR{T{V}_\nu}^2
+\frac{1}{4}\alpha^{(0)}_8g^2\TR{TW_{\mu\nu}}^2
+\frac{i}{2}\alpha^{(0)}_9 g\TR{TW^{\mu\nu}}\TR{T[{V}_\mu,{V}_\nu]}\nonumber\\
&&+\frac{1}{2}\alpha^{(0)}_{10}\TR{T{V}^\mu}^2\TR{T{V}^\nu}^2
+\alpha^{(0)}_{11}g\epsilon^{\mu\nu\rho\lambda}\TR{T{V}_\mu}\TR{{V}_\nu W_{\rho\lambda}}\nonumber\\
&&+\alpha^{(0)}_{12}g\TR{T{V}^\mu}\TR{{V}^\nu W_{\mu\nu}}
+\alpha^{(0)}_{13}gg'\epsilon^{\mu\nu\rho\lambda}B_{\mu\nu}\TR{TW_{\rho\lambda}}\nonumber\\
&&+\alpha^{(0)}_{14}g^2\epsilon^{\mu\nu\rho\lambda}\TR{TW_{\mu\nu}}\TR{TW_{\rho\lambda}}
\end{eqnarray}
Here, a superscript $(i)$ attached to an $\alpha$ coefficients indicates the number of obvious partial derivatives of the Higgs field. As we have mentioned, the coefficients depend on the Higgs field $h$, i.e.,
 \begin{eqnarray}
 \alpha^{(0)}_i(h)=\sum_{j=0}^\infty\alpha^{(0),j}_i\frac{h^j}{\Lambda^j}.
 \end{eqnarray}
For convenience, we omit the functional symbol and denote this as $\alpha^{(0)}_i$.
The contribution $\mathcal{L}_{H}$ with the derivatives of the Higgs field can be written as
\begin{eqnarray}
\mathcal{L}_H &=&(\partial_\mu h)\Big\{\alpha^{(1)}_1\TR{T{V}^\mu}\TR{{V}_\nu{V}^\nu}
    +\alpha^{(1)}_2\TR{T{V}_\nu}\TR{{V}^\mu{V}^\nu}
    +\alpha^{(1)}_3\TR{T{V}_\nu}\TR{T[{V}^\mu,{V}^\nu]}
    \nonumber\\
    &&+i\alpha^{(1)}_4g\TR{T{V}_\nu}\TR{TW^{\mu\nu}}
    +i\alpha^{(1)}_5g'\TR{T{V}_\nu}B^{\mu\nu}
    +i\alpha^{(1)}_6g\TR{T{V}_\nu W^{\mu\nu}}
    +i\alpha^{(1)}_7g\TR{{V}_\nu W^{\mu\nu}}
    \Big\}
    \nonumber\\
    &&+\epsilon^{\mu\nu\rho\sigma}(\partial_\mu h)\Big\{
    \frac{\alpha^{(1)}_{8}}{2}\TR{T{V}_\nu}\TR{T[{V}_\rho,{V}_\sigma]}
    +i\alpha^{(1)}_{9}g\TR{T{V}_\nu}\TR{TW_{\rho\sigma}}
    +i\alpha^{(1)}_{10}g'\TR{T{V}_\nu}B_{\rho\sigma}
    \nonumber\\
    &&+i\alpha^{(1)}_{11}g\TR{T{V}_\nu W_{\rho\sigma}}
    +i\alpha^{(1)}_{12}g\TR{{V}_\nu W_{\rho\sigma}}
    \Big\}
    \nonumber\\
    &&+(\partial_\mu h)(\partial_\nu h)\Big\{
    \alpha^{(2)}_1 \TR{T{V}^\mu}\TR{T{V}^\nu}
    +\alpha^{(2)}_2 \TR{{V}^\mu{V}^\nu}\Big\}
    \nonumber\\
    &&+(\partial_\mu h)^2\Big\{\alpha^{(2)}_3 \TR{T{V}_\nu}^2
    +\alpha^{(2)}_4 \TR{{V}_\nu{V}^\nu}\Big\}
    \nonumber\\
    &&+\alpha^{(3)}_{1} (\partial_\mu h)^2\partial_\nu h\TR{T{V}^\nu}
    +\alpha^{(4)}_{1} (\partial_\mu h)^4
\end{eqnarray}
Note that five new terms corresponding to $\alpha^{(1)}_{j}$ for $j=8..12$ that go beyond the effective Higgs Lagrangian in ref.\cite{LMWangHiggs} have been added, and thus $\mathcal{L}_{H}$ breaks CP.

Aside from the above bosonic contribution, the Lagrangian involving the fermionic interactions can be generalized as
\begin{eqnarray}
    \mathcal{L}_F=-\sum_{f=u,d,l}Y^f\bar{f}_i(g_E^f+g_O^f\gamma_5)fh
\end{eqnarray}
with the (family universal) SM Yukawa coupling $Y^f$.
Here, $g_E^f$ ($g_O^f$) parameterizes the scalar type (pseudo-scalar type) couplings beyond the SM. It is inspired from the ambiguity in the CP property of the light Higgs that comes from the mixing of a CP-even state and a CP-odd state in 2HDM. The SM with a CP-even scalar corresponds to having $g_E^f=1$ and $g_O^f=0$.

%%%%%%%%%%%%%%
The effective Higgs Lagrangian describes nature in which, irrespective of the CP property, there is no new particles with masses between that of the EW gauge bosons and of the Higgs. It is valid for \Higgs found at the LHC.
%%%%
\section{$HVV$ effective vertices and unitarity constraints}\label{sec.HVV}
Starting from the effective Higgs Lagrangian, we can read off the full interaction vertexes. In this section, we focus on the Higgs interaction with EW gauge bosons. From the point of view of experiment, the interactions should include all possible tensor structures which can be measured at colliders. The vertex function $V^{\mu\nu}_{HVV}$ ($V=W, Z$) for $HV_\mu(p_1)V_\nu(p_2)$ is taken
    \begin{eqnarray}
        V^{\mu\nu}_{HVV}=g_{HVV}^{(SM)}\left(a_Vg_{\mu\nu}+b_V\frac{p_\mu p_\nu}{m_V^2}+c_V\epsilon_{\mu\nu\rho\lambda}\frac{k_\rho p_\lambda}{m_V^2}\right)
        \label{HVVvertex}
    \end{eqnarray}
Here, $a_V$ rescales the SM coupling and $b_V$ and $c_V$ are the CP conserving and CP violating couplings, respectively, that depend on momenta. They are related to the effective Lagrangian coefficients by
\begin{eqnarray}
g_{HZZ}^{SM}(a_Z-1)
&=&i\left(g^2\alpha^{(1),0}_{4}-g'^2\alpha^{(1),0}_{5}+\frac{g^2}{2}\alpha^{(1),0}_{7}\right)p^2\nonumber\\
&&-i\frac{s_W^2}{\Lambda}\left(g^2\alpha_8^{(0),1}-2g'^2\alpha_1^{(0),1}\right)q_1\cdot q_2\\
\frac{g_{HZZ}^{SM}}{m_Z^2}b_Z
&=&-i(2g^2\alpha^{(1),0}_{4}-g'^2\alpha^{(1),0}_{5}+g^2\alpha^{(1),0}_{7})\nonumber\\
&&+i\frac{s_W^2}{\Lambda}\left(g^2\alpha_8^{(0),1}-2g'^2\alpha_1^{(0),1}\right)\\
\frac{g_{HZZ}^{SM}}{m_Z^2}c_Z
&=&-i(-2g^2\alpha^{(1),0}_9+2{g'}^2\alpha^{(1),0}_{10}-g^2\alpha^{(1),0}_{12})\nonumber\\
&&+4i\frac{s_W^2}{\Lambda}\left(g^2\alpha_{14}^{(0),1}-g'^2\alpha_{13}^{(0),1}\right)\\
g_{HWW}^{SM}(a_W-1)
&=&-\frac{i}{4}g^2\left(\alpha^{(1),0}_6p\cdot k-\alpha^{(1),0}_7p^2\right)\\
\frac{g_{HWW}^{SM}}{m_W^2}b_W
&=&-\frac{i}{2}g^2\left(\alpha^{(1),0}_6+\alpha^{(1),0}_7\right)\\
\frac{g_{HWW}^{SM}}{m_W^2}c_W
&=&\frac{i}{2}g^2\alpha^{(1),0}_{12}
\end{eqnarray}
with $p=q_1+q_2$ and $k=q_1-q_2$. We find that $a_Z$, $b_Z$, and $c_Z$ stem from not only $\mathcal{L}_H$ but also $\mathcal{L}_B$. The latter is suppressed by the strong coupling scalar $\Lambda$.
%%%%%%%%%

In SM, $WW$ and $ZZ$ scattering amplitudes are quadratic divergence if only the weak boson and photon exchange process are considered. The unitarity limit is at $1.4$TeV. 
If our effective theory  is required to be valid at a higher energy scale $\Lambda$ than $1.4$TeV, \Higgs field must be responsible for unitarity recovery. At present, the fact that no other new particle except 125GeV Higgs has been found just corresponds to the case. 
In general, unitarity constraint refers to set unitarity limit to a variational $\Lambda$. Couplings $a_V,b_V,c_V$ depend on a chosen $\Lambda$. It is complex situation. 
Now, one consider a simply case: a strict unitarity constraint that improve unitarity limit to arbitrary energy. 
Hence, the $HVV$ interaction should be constrained by unitarity limits in the $WW$/$ZZ$ scatterings amplitudes.

The Higgs exchange $WW$ scattering amplitude under effective vertex $V^{\mu\nu}_{HWW}$ in Eq. (\ref{HVVvertex}) is
        \begin{eqnarray}
        i\mathcal{M}^{WW}_H/(g_{HWW}^{(SM)})^2
        &=&\left[a_W(\epsilon_{a}\cdot\epsilon_{b})+b_W\frac{(\epsilon_{a}\cdot p_s)(\epsilon_{b}\cdot p_s)}{m_W^2}\right]
            \frac{i}{p_s^2-m_H^2}
            \nonumber\\
            &&\times\left[a_W(\epsilon_{c}\cdot\epsilon_{d})+b_W\frac{(\epsilon_c\cdot p_s)(\epsilon_d\cdot p_s)}{m_{W}^2}\right]
        \nonumber\\
        &&+\Big[a_W(\epsilon_{a}\cdot\epsilon_{c})+b_W\frac{(\epsilon_{a}\cdot p_t)(\epsilon_{c}\cdot p_t)}{m_W^2}\Big]
            \frac{i}{p_t^2-m_H^2}
            \nonumber\\
            &&\times\Big[a_W(\epsilon_{b}\cdot\epsilon_{d})+b_W\frac{(\epsilon_{b}\cdot p_t)(\epsilon_{d}\cdot p_t)}{m_{W}^2}\Big]
        \nonumber\\
        &&+\Big[a_W(\epsilon_{a}\cdot\epsilon_{d})+b_W\frac{(\epsilon_{a}\cdot p_u)(\epsilon_{d}\cdot p_u)}{m_W^2}\Big]
            \frac{i}{p_u^2-m_H^2}
            \nonumber\\
            &&\times\Big[a_W(\epsilon_{b}\cdot\epsilon_{c})+b_W\frac{(\epsilon_{b}\cdot p_u)(\epsilon_{c}\cdot p_u)}{m_{W}^2}\Big]
    \end{eqnarray}
Here, $p_s=p_a+p_b=p_c+p_d$,~$p_t=p_a-k_c=k_d-p_b$, and $p_u=p_a-k_d=k_c-p_b$.
Notice that the CPV term corresponding to $c_W$ in the vertex (\ref{HVVvertex}) does not contribute to the amplitude because of the fourth order totally anti-symmetric tensor $\epsilon^{\mu\nu\rho\sigma}$. Expanding $i\mathcal{M}^{WW}_H$ in terms of the invariant momentum $s$ ($s$ is equal to $p_s^2$ for s-channel, $p_t^2$ for t-channel and $p_u^2$ for u-channel), we find the leading-order term is of order $s^3$,
    \begin{eqnarray}
    {i\mathcal{M}^{WW}_H=\frac{ib_W^2s^3}{128m_W^8}(g_{HWW}^{(SM)})^2(3+\cos^2\theta)^2+\mathcal{O}(s^2).}
   	\end{eqnarray}
That is, the $p_\mu p_\nu$-type 
coupling
in Eq. (\ref{HVVvertex}) involves a worse case than one in the standard EW theory that yields only $s^2$-order divergence when removing the SM Higgs exchange contribution. To avoid the catastrophe, we must limit the vanishing of $b_W$. The left divergence at order $s^2$ is similar to the SM case. We must set $a_W=1$ to cancel the $s^2$-order divergence to recover unitarity of the $WW$ scattering amplitude.

Similarly, we find the same results in $ZZ$ scattering:
    \begin{itemize}
        \item $c_Z$ gives no contribution to the $ZZ$ scattering amplitude;
        \item $b_Z$ is restricted to zero in order to cancel the $s^3$-order divergence;
        \item $a_Z$ must take unity to cancel the remaining divergence in the $ZZ$ scattering amplitude without Higgs exchange.
    \end{itemize}
%%%%%%%%%%%%%%
\section{CPV effects and LHC phenomenology}\label{sec.phenom}
From Sec.\ref{sec.Lag}, recall the two kinds of CPV effects: one from CPV interactions and the other from the ambiguity in the CP property of the Higgs field. The first does not depend on a CP-even or CP-odd Higgs, but the second does. The main CPV interactions that are related to LHC phenomenology are $HZZ$ and $HWW$ measured in $H\rightarrow WW^*$ and $H\rightarrow ZZ^*$ decays, respectively. They are controlled by $c_Z$ and $c_W$ in Eq.\ref{HVVvertex}. The parameters related to the Higgs CP property are fermion couplings: $g_E^f$ and $g_O^f$ for $f=u,d,l$. A popular idea is that \Higgs is regarded as a mixed state of the light CP-even state $H$ and CP-odd state $A$ in the 2HDM:
	\begin{eqnarray}
	h=\cos\theta H+\sin\theta A
	\end{eqnarray}
with mixing angle $\theta$ \cite{CPmixingHiggs}. The scalar couplings then take the form $g_E^f=\cos\theta$, but the pseudo-scalar couplings still need to be determined by the models. More generally, the fermion coupling of CP-even $H$ is not required, similar to those in SM, which relaxes the limit for $g_E^f$. Thus, there are eight parameters describing CPV effects: $c_Z,~c_W$, $g_E^f$, and $g_O^f$ ($f=u,d,l$). 

To investigate CPV effects in detail, we classify the five classes in terms of two CPV sources: ambiguous CP state and CPV couplings. The former can induce CPV interactions, and the later do not vanish even if $h$ reduces to CP-even eigenstate. Five classes are listed as
        \begin{itemize}
        \item case 0: SM Higgs with $c_Z=c_W=0$, $g_E^f=1$ and $g_O^f=0$---there is no free parameter and no CPV effect;
        \item case 1: pure CP-even Higgs with $g_O^f=0, g^f_E=1$ and $c_Z,c_W$ free---this is an extended SM Higgs with anomalous $HVV$ coupling. CPV effects only come from CP-even Higgs interactions;
        \item case 2: 2HDM Higgs with $c_Z=c_W=0$, $g_E^f=\cos\theta$ and $g_O^f$ free---the dependent parameters are the CP mixing angle $\theta$ and three pseudo-scalar couplings $g^f_O$. The CPV effects complete stem from  CP-mixing state. CPV in the case can be transferred to all Higgs interactions, which controlled by CP-mixing angle;
        \item case 3: multi-CPV Higgs only with $g_E^f=\cos\theta$ and other parameters free---this is a simple combination between ambiguous CP defined state and CPV couplings to EW bosons. CPV interactions are not vanishing even if CP mixing angle tends to zero;
        \item case 4: the most general Higgs with all parameters free---this is an extension of case 3. Ambiguous CP defination is not required to come from CP eigenstate mixing. 
    \end{itemize}
These cases  completely describe all possible CP properties of \Higgs.

Now, let us consider the Higgs decay widths at the LHC. ATLAS and CMS have measured signal strengths at $ZZ^*, WW^*, \gamma\gamma, \tau\tau, b\bar{b}$ decay channels. They are sensitive to different parameters.

Under the narrow width approximation, we get the width of
$\Gamma_{H\rightarrow VV^{(*)}\rightarrow 4l}$(V=Z or W)
\begin{eqnarray}
&&\Gamma_{H \rightarrow VV^{*} \rightarrow f_1f_2f_3f_4}\nonumber\\
&&=\frac{1}{\pi^2}\int dm_1^2 dm_2^2
\frac{m_V\Gamma_{V\rightarrow f_1f_2}m_{V^{*}}\Gamma_{V^{(*)}\rightarrow f_3f_4}
\Gamma_{H\rightarrow VV^{*}}}
{[(m_1^2-m_V^2)^2+m_V^2\Gamma_V^2]
[(m_2^2-m_{V^{*}}^2)^2+m_{V^*}^2\Gamma_{V^{*}}^2]}.
\end{eqnarray}
Here $m_1$ and $m_2$ are the virtualities for the intermediate vector bosons. $\Gamma_V$($\Gamma_{V^{*}}$) and $m_V$($m_{V^{*}}$) are the total width and mass of V(V$^{*}$), respectively. The Higgs decay width to $ZZ^*$ ($WW^*$) with $HZZ$ ($HWW$) couplings given by ref.\cite{HVVexp} is written
    \begin{eqnarray}
        \Gamma_{H \rightarrow VV^*}&=&\frac{G_F\beta}{D\sqrt{2}m_H^3m_V^4\pi}(4a_V^2(\beta^2+12m_1^2m_2^2)m_V^4+b_V^2\beta^4\nonumber\\
        &&+32c_V^2m_1^2m_2^2\beta^2+4a_Vb_Vm_V^2\beta^2\sqrt{\beta^2+4m_1^2m_2^2}),
    \end{eqnarray}
where D is 64(32) for V=Z(W) and the $\beta$ is defined as
    \begin{eqnarray}
        \beta=\sqrt{(m_H^2-(m_1 - m_2)^2)(m_H^2-(m_1 + m_2)^2)}.
    \end{eqnarray}
The sequential decay widths of Z or W into fermions are
    \begin{eqnarray}
        \Gamma_{Z\rightarrow
        f\bar{f}}&=&\frac{G_Fm_Z^3}{6\sqrt{2}\pi}(v_f^2+a_f^2),\\
        \Gamma_{W\rightarrow
        f\bar{\nu_f}}&=&\frac{G_Fm_W^3}{6\sqrt{2}\pi}.
    \end{eqnarray}

Higgs di-photon decay is an important channel appearing at one-loop level. The main contributions come from the $W$ loop and the top loop in the SM. Because there are more tensor structures beyond the SM, the effective $HWW$ coupling may correct the $W$ loop contribution. Its result however shows that the CP violating term with $\epsilon^{\mu\nu\rho\sigma}$ makes no contribution because the Higgs di-photon decay conserves CP. Thus, the $H\rightarrow\gamma\gamma$ amplitude is a sum of SM-like terms with $g_E^f$ rescaling and fermionic pseudo-scalar terms with $g_O^f$,
\begin{eqnarray}
    \Gamma_{\gamma\gamma}&=&\frac{\alpha^2g^2}{4^5\pi^3}
        \Bigg(\left|\frac{4}{3}g^u_EF_{1/2}(\tau_t)+\frac{1}{3}g^d_EF_{1/2}(\tau_b)+F_1(\tau_W)\right|^2
        \\
        &&+\left|\frac{4}{3}g^u_OA_{1/2}(\tau_t)+\frac{1}{3}g^d_OA_{1/2}(\tau_b)\right|^2\Bigg)
\end{eqnarray}
where
    \begin{eqnarray}
    F_1&=&2+3\tau+3\tau(2-\tau)f(\tau)
    \\
    F_{1/2}&=&-2\tau[1+(1-\tau)f(\tau)]
    \\
    A_{1/2}&=&-2\tau f(\tau)
    \end{eqnarray}
and $\tau_i=4m_i^2/m_H^2$ \cite{HiggsHunter}.

The Higgs decay to fermion pairs can be expressed as
\begin{eqnarray}
\Gamma_{f\bar{f}}=\frac{N_CG_F}{4\sqrt{2}\pi} m_f^2m_H
(1-\frac{4m_f^2}{m_H^2})^{\frac{3}{2}}
\left((g_E^f)^2+R^{f}(g_O^f)^2\right)
\end{eqnarray}
with QCD correction factor $R^f$ from the scalar and pseudo-scalar decays \cite{decayQCDcorr}.
%%%%%
\section{Fitted results}\label{sec.fit}
The signal strength of $pp\rightarrow H\rightarrow XX$ at the LHC has been measured by ATLAS and CMS.
In theory, the signal strength can be calculated using the narrow-width approximation
\begin{eqnarray}
\mu_X^{th}=\frac{\sigma^{th}(pp\rightarrow H)}{\sigma^{SM}(pp\rightarrow H)}
\frac{\Gamma^{th}(H\rightarrow XX)}{\Gamma^{SM}(H\rightarrow XX)}.
\end{eqnarray}
However, to calculate Higgs production channels is very complicated. To avoid the problem, we adopt relative ratios of the branching ratios \cite{reRatio}
    \begin{eqnarray}
        D_{X}\equiv \frac{\mu_X}{\mu_{ref}}.
    \end{eqnarray}
with reference $\mu_{ref}$. Thus it only depends on relative decay widths
\begin{eqnarray}
D^{th}_{X}={\frac{\Gamma^{th}(H\rightarrow XX)}{\Gamma^{SM}(H\rightarrow XX)}}
\Bigg/{\frac{\Gamma^{th}_{ref}}{\Gamma^{SM}_{ref}}}.
\end{eqnarray}

However, the signal strength of $b\bar{b}$ channel, $\mu_b$, is measured from the associated production of the higgs boson and the electroweak gauge boson(W or Z)(VH), which differs from $\mu_Z$ and $\mu_W$ that are dominated by the gluon fusion channel (ggF). The production channel will not drop out in the retio of $\mu_b/\mu_{ref}$ if the production modes are different between them. 
An  alternative scheme conveniently adopted by experimental collaborations is to group five production modes, gluon fusion(ggF), vector boson fusion(VBF), associated production with vector bosons(VH), and associated production with top pair(ttH), into two effective modes, ggF$+$ttH and VBF$+$VH. These combined signal strengths are calculated in paper \cite{combinedsingalstrength} and listed in Tab. \ref{tab.signal}. 
    \begin{table}[htdp]
    \caption{Combined signal strength: $\mu^{ggF+ttH}$ and $\mu^{VBF+VH}$ for different decay modes. $VV$ takes $WW, ZZ$.}
    \begin{center}
    \begin{tabular}{c|c|c}
    \hline
    \hline
    Channel & $\mu^{ggF+ttH}$ & $\mu^{VBF+VH}$
    \\
    \hline
    $VV$ & $0.91\pm0.16$ & $1.01\pm0.49$
    \\
    $\gamma\gamma$ & $0.98\pm0.28$ & $1.72\pm0.59$
    \\
    $b\bar{b}$ & $-0.23\pm2.86$ & $0.97\pm0.38$
    \\
    $\tau\tau$ & $1.07\pm0.71$ & $0.94\pm0.65$
    \\
    \hline
    \hline
    \end{tabular}
    \end{center}
    \label{tab.signal}
    \end{table}
Thus, two choices of reference signal strengths can be made independently. One is the sum of all channels from $\mu^{ggF+VF}$, i.e. $\mu_{ref1}=\sum_{X=Z,W,\gamma,b,\tau}\mu^{ggF+VF}_X$; the other is $\mu_{ref2}=\sum_{X=Z,W,\gamma,b,\tau}\mu^{VBF+VH}_X$. 

With the above formula, we can constrain the parameters by minimizing the $\chi^2$ function
    \begin{eqnarray}
        \chi^2=\sum_{X=Z,W,\gamma,b,\tau}\left(\frac{D^{th}_X-D^{exp}_X}{\delta D^{exp}_X}\right)^2.
    \end{eqnarray}

Figures 1--4 show the allowed area for cases 1--4, respectively.
\begin{figure}[htbp]
\begin{center}
   \includegraphics[width=3cm]{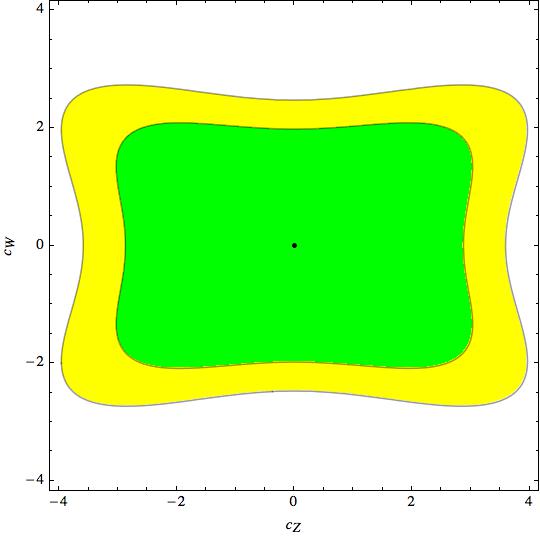}
\caption{Fitted result in $c_Z-c_W$ plane for case 1 at $1\sigma$ (green) and $2\sigma$ (yellow). Black dot represent the best fitted point.}
\label{fig.1}
\end{center}
\end{figure}

\begin{figure}[htbp]
  \centering
\begin{minipage}[c]{0.28\textwidth}
\centering
  \includegraphics[width=3cm]{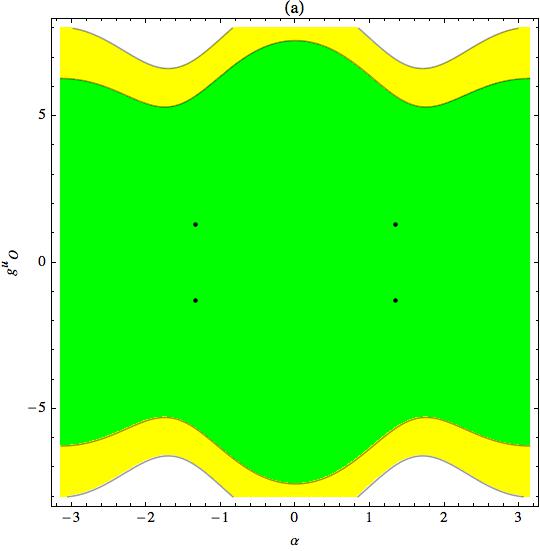}
\end{minipage}%
\begin{minipage}[c]{0.28\textwidth}
\centering
  \includegraphics[width=3cm]{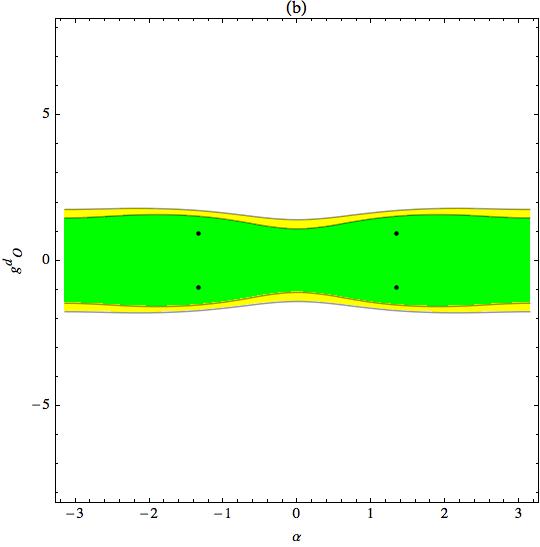}
\end{minipage}%
  \centering
\begin{minipage}[c]{0.28\textwidth}
\centering
  \includegraphics[width=3cm]{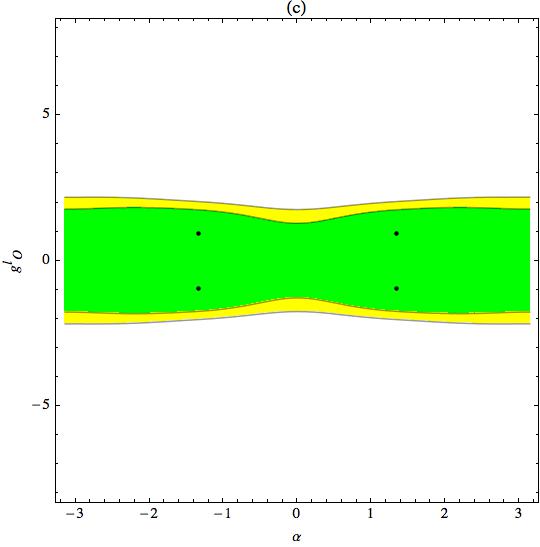}
\end{minipage}
\caption{Fitted results for case 2 at $1\sigma$ (green) and $2\sigma$ (yellow) in the planes of (a) $\alpha-g^u_O$, (b) $\alpha-g^d_O$ and (c) $\alpha-g^l_O$. Other parameters are taken from the best fit values. Black dots represent the best fitted points.}
\label{fig.2}
\end{figure}

\begin{figure}[htbp]
  \centering
\begin{minipage}[c]{0.28\textwidth}
\centering
  \includegraphics[width=3cm]{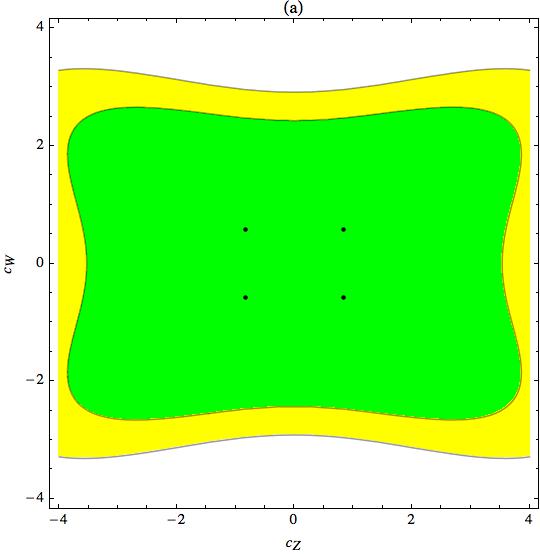}
\end{minipage}%
\begin{minipage}[c]{0.28\textwidth}
\centering
  \includegraphics[width=3cm]{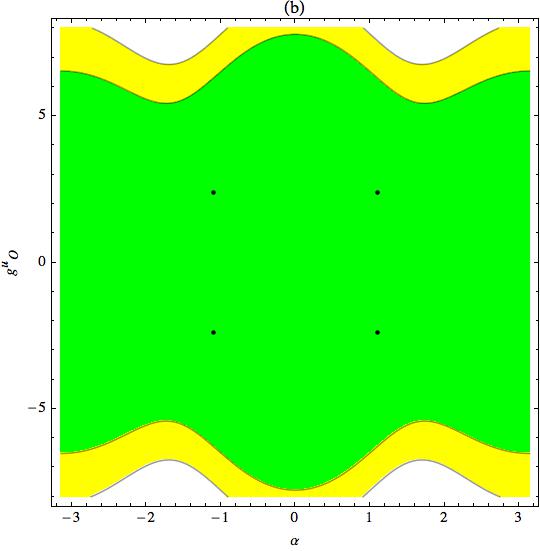}
\end{minipage}
  \centering
\begin{minipage}[c]{0.28\textwidth}
\centering
  \includegraphics[width=3cm]{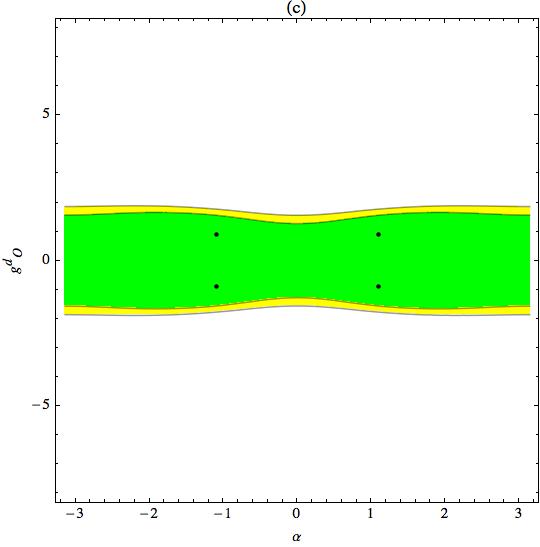}
\end{minipage}%
\begin{minipage}[c]{0.28\textwidth}
\centering
  \includegraphics[width=3cm]{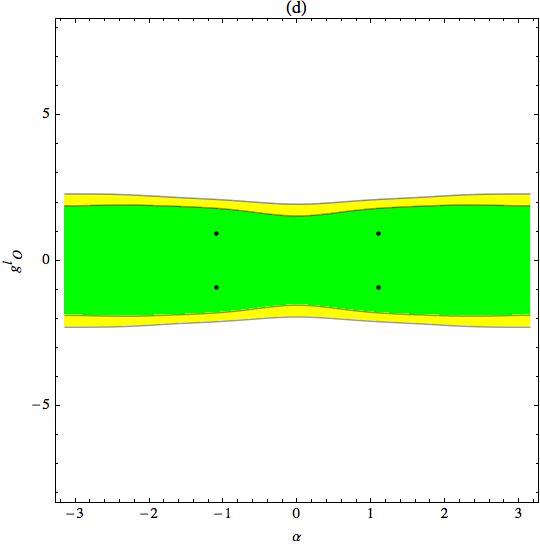}
\end{minipage}
\caption{Fitted results for case 3 at $1\sigma$ (green) and $2\sigma$ (yellow) in the planes of (a) $c_Z-c_W$, (b) $\alpha-g^u_O$, (c) $\alpha-g^d_O$ and (d) $\alpha-g^l_O$. Other parameters are taken from best-fit values. Black dots represent the best fitted points.}
\label{fig.3}
\end{figure}

In Fig.~\ref{fig.1}, $c_Z$ is left more space than $c_W$. The reason is that apart from the $H\rightarrow VV$ decay, $c_W$ also takes part in di-photon decay to constraint it more, but $c_Z$ does not. The same case appears in Fig.~\ref{fig.3}(a).
Case 2 in Fig.~\ref{fig.2} shows that the pseudo-scalar couplings $g^d_O$ and $g^l_O$ have more deviation from SM that $g^u_O$. However, $g^u_O$ has more error.
%An interesting result appears in Fig.~\ref{fig.3}(b) in that the origin which corresponds to SM %is out of the $1\sigma$ allowed area, thus hinting at new physics. 
More details will follow using updated LHC data.

\begin{figure}[htbp]
  \centering
\begin{minipage}[c]{0.28\textwidth}
\centering
  \includegraphics[width=3cm]{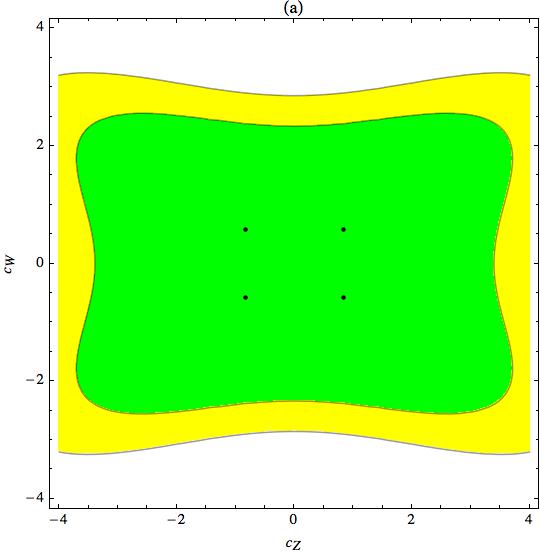}
\end{minipage}%
\begin{minipage}[c]{0.28\textwidth}
\centering
  \includegraphics[width=3cm]{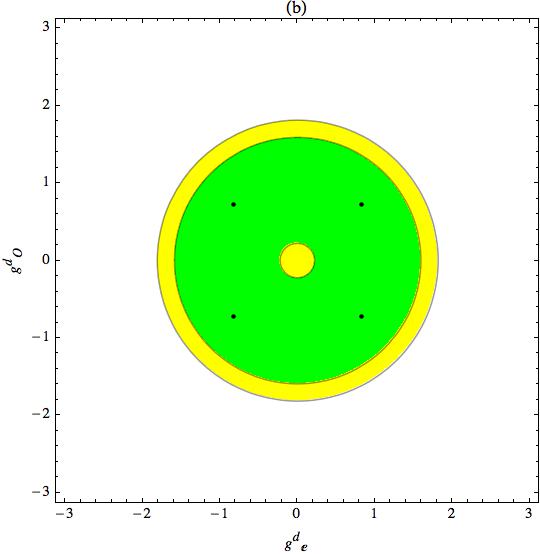}
\end{minipage}%
\begin{minipage}[c]{0.28\textwidth}
\centering
  \includegraphics[width=3cm]{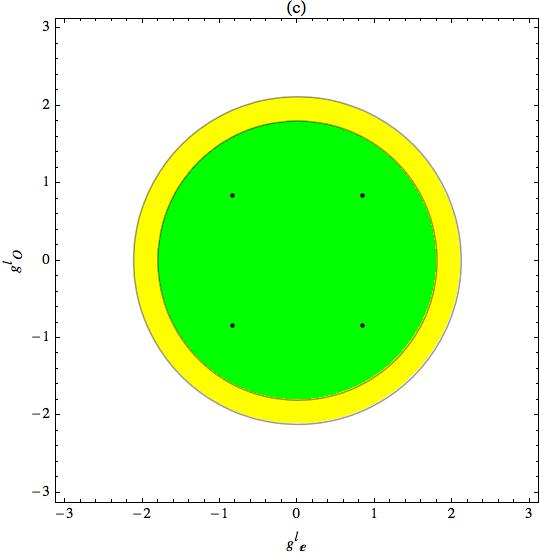}
\end{minipage}
\caption{Fitted results for case 4 at $1\sigma$ (green) and $2\sigma$ (yellow) in the planes of (a) $c_Z-c_W$, (b) $g^d_E-g^d_O$ and (c) $g^l_E-g^l_O$. Other parameters are taken from best-fit values. Black dots present the best fitted points.}
\label{fig.4}
\end{figure}

In case 4, Fig.~\ref{fig.4}(a) shows that $c_Z$ and $c_W$ almost have the same allowed areas as that in case 3. It is not enlarged despite more free parameters. The scalar coupling of the up quark, $g^u_{E}$, is not sensitive to the minimum of chi-squared function, and is not plotted. The pseudo-scalar couplings of the d-quark and leptons, $g^d_{O}$ and $g^l_{O}$, exhibit an observable deviation to SM in Fig.~\ref{fig.4}(b,c).

\section{Summary}\label{sec.summary}
An Higgs effective Lagrangian is completed, which
 provides a complete CPV sources and a general parameter space to describe CP properties of \Higgs and helps to improve the theoretical calculation of the Higgs signal strength.
Using the effective Higgs Lagrangian, we investigated two possible sources of CPV, corresponding to ambiguous CP defined state and CPV interactions.  The unitarity limit of $WW$ and $ZZ$ scatterings only requires $b_V$ to vanish and keeps $c_V$ independent. By fitting signal strengths of the Higgs, the pseudo-scalar coupling $g^f_O$ and anomalous coupling $c_Z$ obviously exhibit deviations to the SM results. The allowed area of $c_Z$ is not noticeably larger when more free parameters are included from cases 1 through 4. All cases have shown that there are a abundant parameter space, even if unitarity limit is severely set to arbitrary large energy. So,
the two kinds of CPV from ambiguous CP definition and anomalous couplings can coexist. 
Even under
These results come from an imprecise signal strength, which will be improved at the updated LHC in the near future.

\section*{acknowledgement}
We thank Prof. Qing Wang for helpful discussions. This work is
supported by the National Natural Science Foundation of China (No.
11105152).

%%%%%%%%%%%%%%%%%


\begin{thebibliography}{1}
\bibitem{HiggsFound}
    G.Aad {\it et al.} [ATLAS Collaboration], Phys. Lett. B 710, 49 (2012);
    S. Chatrchyan {\it et al.} [CMS Collaboration], arXiv:1202.1488 [hep-ex].
\bibitem{2HDM}
    G.C. Branco, P.M. Ferreira, L. Lavoura, M.N. Rebelo, M. Sher, and J.P. Silva, Phys. Rep. 516, 1 (2012).
\bibitem{HiggsCP}
    A. Freitas and P. Schwaller, Phys. Rev. D 87, 055015(2013);
    J. Chang, K. Cheung, P.-Y. Tseng, T.-C. Yuan, Phys. Rev. D 87, 035008 (2013).
\bibitem{reRatio}
    A. Djouadi, G. Moreau, arXiv:1303.6591[hep-ph].
\bibitem{HVVexp}
    R.M. Godbole, D.J. Miller, M.M. Muhlleitner, JHEP 0712 (2007) 031.
\bibitem{ChangPRD}
    W.-F. Chang, W.-P. Pan, F. Xu, Phys. Rev. D88, 033004(2013).
\bibitem{GavelaJHEP}
	M.B. Gavela, J. Gonzalez-Fraile, M.C. Gonzalez-Garcia, L. Merlo, S. Rigolin, J. Yepes, JHEP  1410 (2014), 044.
\bibitem{LMWangHiggs}
    Li-Ming Wang, Qing Wang, arXiv:hep-ph/0605104.
\bibitem{AppelquistEWCL}
    T. Appelquist, G.-H. Wu, Phys. Rev. D 48, 3235 (1993).
\bibitem{ATLAS2013}
    The ATLAS Collaboration, ATLAS-CONF-2013-014.
\bibitem{CMS2013}
    The CMS Collaboration, CMS PAS HIG-13-005.
\bibitem{CPmixingHiggs}
    N. Oshimo, Prog. Theor. Exp. Phys. (2013) 083B04.
\bibitem{HiggsHunter}
    J.F. Gunion, H.E. Haber, G. Kane, S. Dawson, {\it The Higgs Hunter's Guide}, Perseus Publishing, 2000;
    J. Ellis, M. K. Gaillard, and D. V. Nanopoulos, Nucl. Phys. B106, 292 (1976); A. Djouadi, M. Spira, and P. M. Zerwas, Phys. Lett. B 264, 440 (1991); M. Spira, A. Djouadi, D. Graudenz, and P. M. Zerwas, Nucl. Phys. B453, 17 (1995).
\bibitem{decayQCDcorr}
    M. Spira, Fortschr.Phys. 46(1998) 203.
\bibitem{combinedsingalstrength}
	G. Belanger, B. Dumont, U. Ellwanger, J.F. Gunion, S. Kraml, Phys. Rev. D88, 075008 (2013).
\end{thebibliography}
\end{document}